\documentclass{midl} 

\usepackage{mwe} 

\jmlrproceedings{MIDL}{Medical Imaging with Deep Learning}
\jmlrpages{}

\title[Visual explanation]{Visualizing chest X-ray dataset biases using GANs}

 \midlauthor{\Name{Hao Liang} \Email{hl106@rice.edu}\\
  \Name{Kangqi Ni} \Email{kn22@rice.edu}\\
  \Name{Guha Balakrishnan} \Email{guha@rice.edu}\\
  \addr Department of Electrical and Computer Engineering, Rice University, USA}

\begin{document}

\maketitle
\vspace{-3.5em}
\begin{abstract}
Recent work demonstrates that images from various chest X-ray datasets contain visual features that are strongly correlated with protected demographic attributes like race and gender. This finding raises issues of fairness, since some of these factors may be used by downstream algorithms for clinical predictions. In this work, we propose a framework, using generative adversarial networks (GANs), to visualize what features are most different between X-rays belonging to two demographic subgroups.
\end{abstract}

\begin{keywords}
Chest X-rays, fairness, bias, explainability, generative adversarial networks (GANs)
\end{keywords}

\section{Introduction}
Recent studies have demonstrated that patient bio-information like age, race, and gender are predictable from chest X-ray (CXR) images alone using deep learning models\cite{gichoya2022ai,karargyris2019age,duffy2022confounders}. For example, in the ``Reading Race'' study, deep classifiers trained to predict race achieve $0.99$ AUROC on several CXR datasets~\cite{gichoya2022ai}. This finding raises the question: ``What visual cues discriminate different races?'' Answering such a question can help mitigate potentially biased behavior of downstream algorithms that make decisions using this data. 
In this work, we propose a framework to visually explain the principal differences between different demographic subgroups in a medical imaging dataset. We first train an unconditional generative adversarial network (GAN) \cite{goodfellow2020generative,liang2020controlled,lin2022raregan} on the given image dataset. Next, we project the images onto the (trained) GAN's latent space and compute a direction in the latent space that differentiates a pair of classes (e.g., ``Black'' vs. ``White'' race groups). We traverse the latent space along that direction to produce image sequences that depict the main morphological and appearance changes in moving from one class to another.

There are related works that focus on visualizing subgroup differences associated with clinical attributes. One such study uses autoencoders~\cite{cohen2021gifsplanation}, which often produce blurry samples that do not clearly capture structural information. Others train \emph{conditional} versions of GANs~\cite{singla2023explaining,dravid2022medxgan}, an expensive process since the GAN must be trained from scratch for each attribute of interest. In contrast to all these works, we demonstrate that deep generative models may be a useful tool to the medical imaging community to understand the biases within a medical imaging dataset. 

\begin{figure}[t!]
    \centering
    {\label{fig:1}\includegraphics[width=0.8\textwidth]{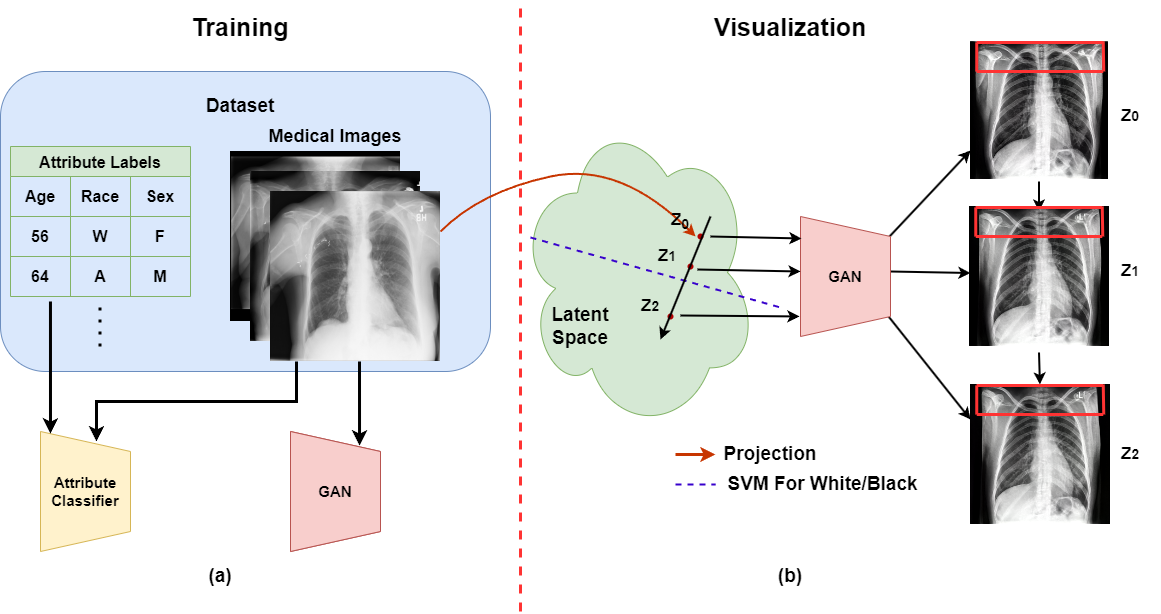}}
     \caption{\textbf{Framework of our proposed method.} (a) We train a GAN on an image dataset, and a binary classifier on the images and labels for a demographic prediction task (e.g., White vs. Black race). (b) We project a subset of images onto the trained GAN's latent space. To ensure the projected images are reasonably reconstructed, we only keep projected images whose labels (predicted by the attribute classifier trained in (a)) agree with their original labels. We also fit an SVM hyperplane to separate the two classes in the latent space. Finally, we visualize the differences between the classes by starting at a latent code corresponding to a random image, and traversing along the normal direction of the SVM hyperplane, to generate a sequence of images showing a transformation.}
    \label{fig:perf}
\vspace{-2em}
\end{figure}


\section{Method}
\label{sec:method}
Our method consists of several components, visualized in Fig.~\ref{fig:1} and described below.

\textbf{Generator training:}
We train an unconditional StyleGAN2 generator~\cite{karras2020training} $G(\cdot): R^d \rightarrow R^{H \times W \times 1}$, following the default training procedure introduced in that paper. $d$ is the dimension of the ``latent space'' of the generator, and $H$ and $W$ are the height and width of the generated CXR. In our experiments, we trained $G(\cdot)$ on Chexpert~\cite{irvin2019chexpert}, a large public dataset containing $224,316$ CXRs. We only used frontal views, yielding $164,548$ CXRs. The training procedure takes roughly 24 hours on two Nvidia A100 GPUs.

\textbf{Attribute classifier training:} We train a separate deep attribute classifier $C(\cdot): R^{H \times W \times 1} \rightarrow R^1$ for each per-image binary attribute provided in the dataset. For multi-class labels such as race, we train a separate binary classifier for each pair of races.

\textbf{Image projection/SVM training:} Next, we follow the process introduced in \cite{karras2020analyzing}  to project a subset of CXR images $\{X_i\}_{i=1}^N$ onto $G$'s latent space, yielding latent codes $\{z_i\}_{i=1}^N$. We only retain those projected images whose labels (predicted by $C$) are the same as the original labels $\{L_i\}_{i=1}^N$, i.e., $C(G(z_i)) = L_i$. We then train a linear SVM to predict $L_i$ from $z_i$.

\textbf{Image sequence generation:} The normal vector $v$ of the trained SVM's hyperplane identifies the direction that best differentiates the two classes. We will use this fact to generate image sequences depicting the principal perceptual changes needed to convert a CXR belonging to one demographic class to another. In particular, we select the latent vector corresponding to a random dataset CXR, and move towards the opposite class in latent space in the direction of $v$. We concatenate images generated by intermediate latent codes along this traversal to produce a sequence.

\begin{figure}[t!]
    \centering
    {\label{fig:1}\includegraphics[width=0.9\textwidth]{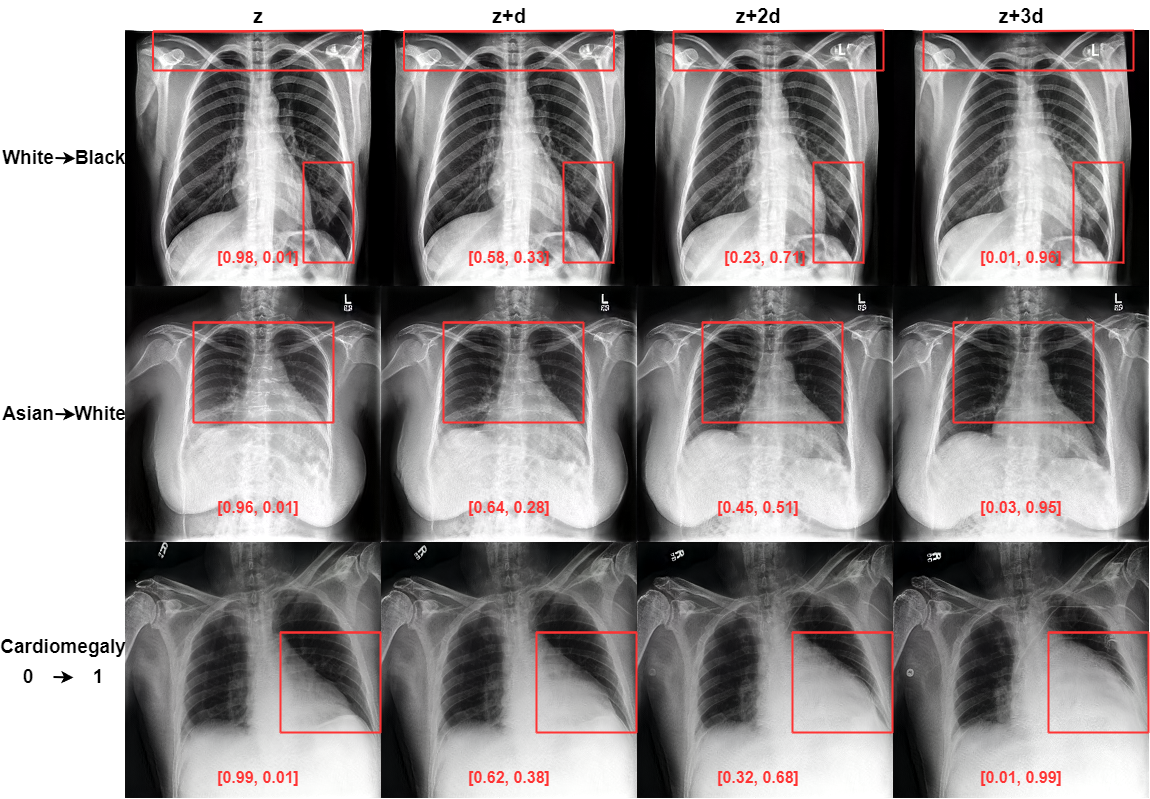 }}
    \caption{\textbf{Sample visualization results.} The left column corresponds to the projected initial image and the last three columns show images generated at different traversal distances in the latent space. The red text indicates the output probabilities predicted by the attribute classifier for each class. For example, the top left [0.98, 0.01] indicate the CXR has a $98\%$ possibility of being white and $1\%$ possibility of being black. We also use red boxes to highlight the areas that visually vary the most. For White/Black, the shoulder bone and right lung structures change shape, and the lungs become more opaque. For Asian/White, the entire chest shape changes and grows larger. These visualizations also explain why the Reading Race study~\cite{gichoya2022ai} did not find race prediction to significantly change when blocking local regions. The proposed applied to \textit{Cardiomegaly} enlarges the heart, in agreement with the known effect of that disease.}
    \label{fig:res}
\vspace{-1.5em}
\end{figure}

\section{Results and discussion}
We demonstrate our framework on ChexPert with \textit{race} as the target attribute. We also validate our approach on the clinical attribute \textit{Cardiomegaly}, which induces a known physiological change (enlarged heart). Sample results are shown and explained in Fig.~\ref{fig:res}.

{\bf Conclusion} Our results show that an unconditional generative adversarial network can be a useful tool for visualizing differences between demographic groups of a CXR dataset. Our framework is fast and flexible, and can be applied to any binary attribute labels in the dataset. Future work includes analyzing generated sequences to thoroughly investigate demographic differences, and comparing results across different generative models.


\clearpage
\bibliography{midl-samplebibliography}

\begin{thebibliography}{12}
\providecommand{\natexlab}[1]{#1}
\providecommand{\url}[1]{\texttt{#1}}
\expandafter\ifx\csname urlstyle\endcsname\relax
  \providecommand{\doi}[1]{doi: #1}\else
  \providecommand{\doi}{doi: \begingroup \urlstyle{rm}\Url}\fi

\bibitem[Cohen et~al.(2021)Cohen, Brooks, En, Zucker, Pareek, Lungren, and
  Chaudhari]{cohen2021gifsplanation}
Joseph~Paul Cohen, Rupert Brooks, Sovann En, Evan Zucker, Anuj Pareek,
  Matthew~P Lungren, and Akshay Chaudhari.
\newblock Gifsplanation via latent shift: a simple autoencoder approach to
  counterfactual generation for chest x-rays.
\newblock In \emph{Medical Imaging with Deep Learning}, pages 74--104. PMLR,
  2021.

\bibitem[Dravid et~al.(2022)Dravid, Schiffers, Gong, and
  Katsaggelos]{dravid2022medxgan}
Amil Dravid, Florian Schiffers, Boqing Gong, and Aggelos~K Katsaggelos.
\newblock medxgan: Visual explanations for medical classifiers through a
  generative latent space.
\newblock In \emph{Proceedings of the IEEE/CVF Conference on Computer Vision
  and Pattern Recognition}, pages 2936--2945, 2022.

\bibitem[Duffy et~al.(2022)Duffy, Clarke, Christensen, He, Yuan, Cheng, and
  Ouyang]{duffy2022confounders}
Grant Duffy, Shoa~L Clarke, Matthew Christensen, Bryan He, Neal Yuan, Susan
  Cheng, and David Ouyang.
\newblock Confounders mediate ai prediction of demographics in medical imaging.
\newblock \emph{npj Digital Medicine}, 5\penalty0 (1):\penalty0 188, 2022.

\bibitem[Gichoya et~al.(2022)Gichoya, Banerjee, Bhimireddy, Burns, Celi, Chen,
  Correa, Dullerud, Ghassemi, Huang, et~al.]{gichoya2022ai}
Judy~Wawira Gichoya, Imon Banerjee, Ananth~Reddy Bhimireddy, John~L Burns,
  Leo~Anthony Celi, Li-Ching Chen, Ramon Correa, Natalie Dullerud, Marzyeh
  Ghassemi, Shih-Cheng Huang, et~al.
\newblock Ai recognition of patient race in medical imaging: a modelling study.
\newblock \emph{The Lancet Digital Health}, 4\penalty0 (6):\penalty0
  e406--e414, 2022.

\bibitem[Goodfellow et~al.(2020)Goodfellow, Pouget-Abadie, Mirza, Xu,
  Warde-Farley, Ozair, Courville, and Bengio]{goodfellow2020generative}
Ian Goodfellow, Jean Pouget-Abadie, Mehdi Mirza, Bing Xu, David Warde-Farley,
  Sherjil Ozair, Aaron Courville, and Yoshua Bengio.
\newblock Generative adversarial networks.
\newblock \emph{Communications of the ACM}, 63\penalty0 (11):\penalty0
  139--144, 2020.

\bibitem[Irvin et~al.(2019)Irvin, Rajpurkar, Ko, Yu, Ciurea-Ilcus, Chute,
  Marklund, Haghgoo, Ball, Shpanskaya, et~al.]{irvin2019chexpert}
Jeremy Irvin, Pranav Rajpurkar, Michael Ko, Yifan Yu, Silviana Ciurea-Ilcus,
  Chris Chute, Henrik Marklund, Behzad Haghgoo, Robyn Ball, Katie Shpanskaya,
  et~al.
\newblock Chexpert: A large chest radiograph dataset with uncertainty labels
  and expert comparison.
\newblock In \emph{Proceedings of the AAAI conference on artificial
  intelligence}, volume~33, pages 590--597, 2019.

\bibitem[Karargyris et~al.(2019)Karargyris, Kashyap, Wu, Sharma, Moradi, and
  Syeda-Mahmood]{karargyris2019age}
Alexandros Karargyris, Satyananda Kashyap, Joy~T Wu, Arjun Sharma, Mehdi
  Moradi, and Tanveer Syeda-Mahmood.
\newblock Age prediction using a large chest x-ray dataset.
\newblock In \emph{Medical Imaging 2019: Computer-Aided Diagnosis}, volume
  10950, pages 468--476. SPIE, 2019.

\bibitem[Karras et~al.(2020{\natexlab{a}})Karras, Aittala, Hellsten, Laine,
  Lehtinen, and Aila]{karras2020training}
Tero Karras, Miika Aittala, Janne Hellsten, Samuli Laine, Jaakko Lehtinen, and
  Timo Aila.
\newblock Training generative adversarial networks with limited data.
\newblock \emph{Advances in neural information processing systems},
  33:\penalty0 12104--12114, 2020{\natexlab{a}}.

\bibitem[Karras et~al.(2020{\natexlab{b}})Karras, Laine, Aittala, Hellsten,
  Lehtinen, and Aila]{karras2020analyzing}
Tero Karras, Samuli Laine, Miika Aittala, Janne Hellsten, Jaakko Lehtinen, and
  Timo Aila.
\newblock Analyzing and improving the image quality of stylegan.
\newblock In \emph{Proceedings of the IEEE/CVF conference on computer vision
  and pattern recognition}, pages 8110--8119, 2020{\natexlab{b}}.

\bibitem[Liang et~al.(2020)Liang, Yu, Xu, Raj, and Singh]{liang2020controlled}
Hao Liang, Lulan Yu, Guikang Xu, Bhiksha Raj, and Rita Singh.
\newblock Controlled autoencoders to generate faces from voices.
\newblock In \emph{Advances in Visual Computing: 15th International Symposium,
  ISVC 2020, San Diego, CA, USA, October 5--7, 2020, Proceedings, Part I 15},
  pages 476--487. Springer, 2020.

\bibitem[Lin et~al.(2022)Lin, Liang, Fanti, and Sekar]{lin2022raregan}
Zinan Lin, Hao Liang, Giulia Fanti, and Vyas Sekar.
\newblock Raregan: Generating samples for rare classes.
\newblock In \emph{Proceedings of the AAAI Conference on Artificial
  Intelligence}, volume~36, pages 7506--7515, 2022.

\bibitem[Singla et~al.(2023)Singla, Eslami, Pollack, Wallace, and
  Batmanghelich]{singla2023explaining}
Sumedha Singla, Motahhare Eslami, Brian Pollack, Stephen Wallace, and Kayhan
  Batmanghelich.
\newblock Explaining the black-box smoothly—a counterfactual approach.
\newblock \emph{Medical Image Analysis}, 84:\penalty0 102721, 2023.

\end{thebibliography}

\end{document}